\documentclass[%
 aip,
 amsmath,amssymb, dvipsnames,
 reprint,%
]{revtex4-1}

\usepackage{graphicx}
\usepackage{dcolumn}
\usepackage{bm}
\usepackage{setspace}
\usepackage{lipsum}
\usepackage{mathtools,xcolor}
\usepackage{newtxtext}

\usepackage{hyperref}
\hypersetup{
	colorlinks,
	linkcolor={red!60!black},
	citecolor={red!60!black},
	urlcolor={red!80!black}
}

\begin{document}

\title{Moment-based parameter inference with error guarantees for stochastic reaction networks}

\author{Zekai Li}
 \altaffiliation[Electronic mail: ]{zekai.li18@imperial.ac.uk}
\author{Mauricio Barahona}
 \author{Philipp Thomas}
 \altaffiliation[Corresponding author: ]{p.thomas@imperial.ac.uk}
\affiliation{ 
Department of Mathematics, Imperial College London, London SW7 2AZ, United Kingdom
}

\date{\today}

\begin{abstract} 
Inferring parameters of models of biochemical kinetics from single-cell data remains challenging because of the uncertainty arising from the intractability of the likelihood function of stochastic reaction networks. Such uncertainty falls beyond current error quantification measures, which focus on the effects of finite sample size and identifiability but lack theoretical guarantees when likelihood approximations are needed. Here, we propose a method for the inference of parameters of stochastic reaction networks that works for both steady-state and time-resolved data and is applicable to networks with non-linear and rational propensities.  Our approach provides bounds on the parameters via convex optimisation over sets constrained by moment equations and moment matrices by taking observations to form moment intervals, which are then used to constrain parameters through convex sets. The bounds on the parameters contain the true parameters under the condition that the moment intervals contain the true moments, thus providing uncertainty quantification and error guarantees. Our approach does not need to predict moments and distributions for given parameters (i.e., it avoids solving or simulating the forward problem), and hence circumvents intractable likelihood computations or computationally expensive simulations. We demonstrate its use for uncertainty quantification, data integration and prediction of latent species statistics through synthetic data from common non-linear biochemical models including the Schl\"ogl model and the toggle switch, a model of post-transcriptional regulation at steady state, and a birth-death model with time-dependent data.
\end{abstract}

\maketitle

\section{Introduction}

Biological processes in single cells are influenced by fluctuations in the timing of biochemical reactions \cite{mcadams1997stochastic,raser2005noise,taniguchi2010quantifying}. Single-cell analysis has become an essential tool to quantify variability across living cells and holds the promise to unravel the mechanisms underlying cellular functions through advances in flow cytometry, microscopy, and omics approaches. Yet the quantitative understanding of these experiments necessitates mechanistic models that account for the observed cell-to-cell variability\cite{shahrezaei2008analytical,thomas2014phenotypic,dattani2017stochastic,gorin2023studying}. The chemical master equation (CME) provides such a mathematical description governing the probability distributions of stochastic reaction networks\cite{gillespie1977exact,goutsias2013markovian,anderson2015stochastic,schnoerr2017approximation,wilkinson2018stochastic,kuntz2021stationary}. Making accurate predictions using these models requires parameter inference from experimental data and quantifying the uncertainty in these predictions, a challenging task prompting more efficient and precise inference methods\cite{schnoerr2017approximation,loskot2019comprehensive,warne2019simulation}. 

Common methods for parameter inference include likelihood-based approaches. Maximum-likelihood methods optimise the probability of observing the data given a model over a set of parameters, whereas Bayesian inference methods focus on the posterior distribution of the parameters and naturally provide uncertainty measures \citep{golightly2006bayesian,boys2008bayesian, jiang2022identification}. Likelihood-based methods have been successfully implemented for ODEs \citep{frohlich2017scalable} and many stochastic gene expression models \citep{suter2011mammalian,zechner2014scalable,lin2019exact, luo2022bisc, kilic2023gene}. The advantage of these methods is that they can, in principle, provide error estimates of parameters such as confidence intervals of maximum likelihood parameter estimates, profile likelihoods, or the credible intervals of the posterior distributions \cite{joshi2006exploiting,kreutz2012likelihood,mitra2019parameter,villaverde2022assessment}.

For stochastic reaction networks, however, the computation of likelihoods relies on explicit solutions of the CME, which quickly become intractable even for relatively simple networks \citep{drovandi2011approximate, wilkinson2013approximate}. Even in the few analytically tractable cases, optimising likelihoods is a tedious, non-convex problem with no guarantees of success even with Bayesian sampling methods. Such difficulties persist when considering inference using moments derived from the underlying probability distribution since moments form an infinite hierarchy of coupled equations, called the \textit{moment equations}, which is impossible to solve in general \citep{grima2012study}. Many authors, therefore, resort to approximations such as moment-closure approximations or system size expansion for inference \citep{grima2012study,luck2016generalized,frohlich2016inference}. 

Likelihood-free inference circumvents these analytical expressions, and a wide variety of approaches are available, including approximate likelihood methods \citep{golightly2005bayesian,tian2007simulated, golightly2011bayesian,andreychenko2012approximate,ocal2022inference}, approximate Bayesian computations  \citep{toni2010simulation,liepe2010abc, liepe2014framework,schalte2020efficient,tankhilevich2020gpabc} and, more recently, machine learning approaches \citep{cranmer2020frontier, jorgensen2022efficient, sukys2022approximating}. Although some of these approximate inference methods provide uncertainty measures analogue to their exact counterparts, e.g., Fisher information\citep{komorowski2011sensitivity} or posterior distributions \citep{fearnhead2014inference}, the underlying approximations and assumptions of asymptotic normality of the estimators introduce additional and difficult to control errors that have to be evaluated on a case-by-case basis. The approximate uncertainty measures may not properly capture the actual variability within the data, resulting in less robust or potentially misleading conclusions \citep{komorowski2011sensitivity}. It thus remains an open question how to design inference methods with error guarantees that can ensure the robustness of their predictions. 

In recent years, a few authors have obtained theoretically guaranteed bounds on the stationary moments \citep{kuntz2016bounding, sakurai2017convex, ghusinga2017exact, dowdy2018bounds, dowdy2018dynamic, sakurai2018optimization,  kuntz2019bounding,hori2020modeling, sakurai2022interval} and the transient moments \citep{dowdy2018dynamic,sakurai2018bounding,holtorf2023tighter} of stochastic reaction networks. These approaches rely on convex optimisation of sets constrained by moment equations, a set of equations involving the moments and the reaction rate parameters, and positive semidefinite constraints on the moments. By optimising over these convex sets with a given set of rate parameters, one can obtain upper and lower bounds on the moments. These bounds provide error guarantees for predicting moments, unlike approximation methods based on system size expansion or moment closure. Whether similar approaches could be utilised to provide error bounds for parameter inference has remained unexplored heretofore.

\begin{figure}[htb!] 
\centering 
\includegraphics[width=\columnwidth]{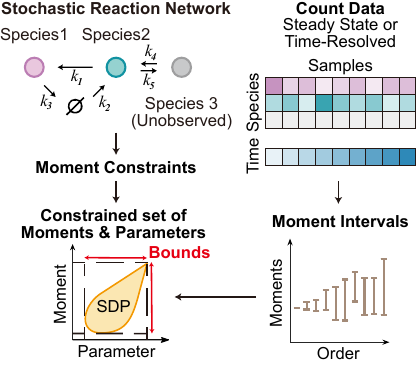}
\caption{\textbf{Flowchart of inference using moment constraints.} Given a stochastic reaction network model (potentially with unobserved species), we use a series of constraints involving the moments and the reaction rate parameters to form a constrained set on them. The mathematical expressions of these constraints can be found in Sec. \ref{sec:mmteqns}. From count data at steady state or time-resolved data, we obtain bootstrap intervals on the moments and input these into the set. By introducing additional variables to replace the non-linear terms in the set, we have a semi-definite program (SDP) over which we optimise with respect to a parameter (or an unobserved moment) to obtain upper and lower bounds.}
\label{fig:Flow1}
\end{figure}

In this work, we present an approach to inference in stochastic reaction networks with unobserved species, which provides upper and lower bounds on parameters. 
Our approach can be applied
to data observed at steady state as well as time trajectory data (i.e. transient data). It takes intervals of the moments as inputs and constrains the parameter space by considering the moment equations and moment matrix constraints to formulate an optimisation problem. The optimisation is turned into a convex semidefinite program (SDP) through a hierarchy of outer approximations with a unique global optimum that allows efficient computation using available SDP solvers. If the input moment intervals contain the true moments, our approach guarantees to provide bounds on the parameters that include the true values. Our method extends to bound moments of unobserved species. A visual summary of our approach is shown in Fig.~\ref{fig:Flow1}. 

The outline of the paper is as follows. In Sec.~\ref{sec:Pre}, we briefly introduce our notation and key facts on the CME and the moment equations in matrix form. We then present in Sec.~\ref{sec:Method} our approach to forming convex-constrained sets in terms of the parameters based on moment equations and moment matrices in both the complete and partial data cases. In Sec.~\ref{sec:Results}, we demonstrate the application of our methods to three biochemical reaction networks with synthetic data: the Schl\"ogl model with fully observed data; a toggle switch to integrate several datasets from different conditions; and a post-transcriptional regulation model with unobserved species. We conclude in Sec.~\ref{Sec:discussion} with a discussion.

\section{Notation and background: The CME and the Moment Equations} 

\label{sec:Pre}

We consider a generic stochastic reaction network as a system with $R$ reactions involving $S$ species:
\begin{equation} \label{eqn:network}
\text{Reaction $r$:}\quad v^{-}_{r1}X_1 + \ldots + v^{-}_{rS}X_S \xrightarrow{a_r} v^{+}_{r1}X_1 + \ldots + v^{+}_{rS}X_S,
\end{equation} 
for $r = 1,\ldots,R$.  Here $v^{\pm}_{rs}$ are the \textit{stoichiometric coefficients} in reaction $r$, with $s=1, \ldots, S $, and  $a_r(\boldsymbol{x})$ is the \textit{propensity function} of reaction $r$ that depends on the state of the system given by the vector of abundances of species 
$\boldsymbol{x}=(x_1,\ldots, x_S) \in \mathcal{X}$ where $\mathcal{X} \subseteq \mathbb{N}^S$ denotes the \textit{state space}, i.e., the set of states that are reachable given any initial state. 

\subsection{Chemical Master Equation}

The CME consists of a set of differential equations of the probability distribution $\boldsymbol{P}_t\coloneqq(P_t(\boldsymbol{x}))_{\boldsymbol{x}\in\mathcal{X}}$ over time $t$ and is commonly formulated in the form
\begin{eqnarray} \label{eqn:CME}
    &&\frac{\mathrm{d} P_t(\boldsymbol{x})}{\mathrm{d} t} = \boldsymbol{P}_tQ(\boldsymbol{x}) \coloneqq \sum_{\boldsymbol{y}\in\mathcal{X}} P_t(\boldsymbol{y})q(\boldsymbol{y},\boldsymbol{x}),\\
    & &q(\boldsymbol{y},\boldsymbol{x}) \coloneqq \sum_{r=1}^{R} a_r(\boldsymbol{y}) (\mathbb{I}_{\boldsymbol{y}+\boldsymbol{v}_r}(\boldsymbol{x})-\mathbb{I}_{\boldsymbol{y}}(\boldsymbol{x})),
\end{eqnarray}
where $\boldsymbol{v}_r := (v^{+}_{r1}-v^{-}_{r1},\ldots,v^{+}_{rS}-v^{-}_{rS})$ denotes the net-stoichiometry of the reaction $r$, and $\mathbb{I}$ is the indicator function defined as
\begin{equation} \label{eqn:indicator}
    \mathbb{I}_{\boldsymbol{x}}(\boldsymbol{y}) \coloneqq \left\{ \begin{array}{ll}
1, & \text{if} \ \boldsymbol{x} = \boldsymbol{y} \\
0, & \text{otherwise.}
\end{array}\right.
\end{equation}

We consider two types of propensity functions: polynomial and rational functions. For polynomial propensities, we assume the functions are of the form
\begin{equation} \label{eqn:polyprop}
    a_r(\boldsymbol{x}) = k_r b_r(\boldsymbol{x}), \quad r = 1,\ldots,R
\end{equation}
where $k_r$ is the rate constant and $b_r(\boldsymbol{x})$ are multivariate polynomials in $\boldsymbol{x}$. For rational propensities\citep{kuntz2019bounding}, we can always rewrite them in the form 
\begin{equation} \label{eqn:rationalprop}
    a_r(\boldsymbol{x}) = k_r \frac{b_r(\boldsymbol{x})}{h(\boldsymbol{x})}, \quad r = 1,\ldots,R,
\end{equation}
where $h$ is the common denominator in all reactions. More specifically, we denote the highest degree of $b_r$ by $d_b$ and the degree of $h$ by $d_h$. Then $h$ can be written as
\begin{equation} 
    h(\boldsymbol{x}) = \sum_{|\boldsymbol{\alpha}|\leq d_h}h_{\boldsymbol{\alpha}} \boldsymbol{x}^{\boldsymbol{\alpha}},
\end{equation}
with $h(\boldsymbol{x}) > 0, \  \forall \boldsymbol{x} \in \mathcal{X}$, and
\begin{equation} 
    b_r(\boldsymbol{x}) = \sum_{|\boldsymbol{\alpha}|\leq d_b}b_{r,{\boldsymbol{\alpha}}} \boldsymbol{x}^{\boldsymbol{\alpha}},\ r = 1,\ldots,R,
\end{equation}
where we introduced the multi-index notation and its order:
\begin{equation} \label{eqn:multiindex}
	\boldsymbol{x}^{\boldsymbol{\alpha}} \coloneqq x^{\alpha_1}_1 x^{\alpha_2}_2 \cdots x^{\alpha_S}_S\quad \text{and} \quad |{\boldsymbol{\alpha}}| \coloneqq \sum_{s=1}^S \alpha_s,
\end{equation}
for a multi-index ${\boldsymbol{\alpha}} = (\alpha_1, \alpha_2, \ldots, \alpha_S)$. We denote the coefficients in $h(\boldsymbol{x})$ by $\boldsymbol{h} \coloneqq (h_{\boldsymbol{\alpha}})_{|\boldsymbol{\alpha}|\leq d_h}$.
Since the polynomial propensities are a special case of rational ones with $h \equiv 1$, we will only discuss the rational ones from now on.

\subsection{Moment Equations}
For a probability distribution $P_t$, we denote the expectation of a function $f\colon\mathcal{X}\rightarrow\mathbb{R}$ as
$\langle f \rangle_{P_t} \coloneqq \sum_{\boldsymbol{x}\in\mathcal{X}}f(\boldsymbol{x})P_t(\boldsymbol{x})$. A \textit{raw moment} is then given by
\begin{align}
\mu^{\boldsymbol{\alpha}}(t)\coloneqq
\left\langle
\boldsymbol{x}^{\boldsymbol{\alpha}} \right\rangle_{P_t}
\end{align}
and the corresponding rational moment
\begin{equation}
y^{\boldsymbol{\alpha}}(t)\coloneqq
\left\langle
\frac{\boldsymbol{x}^{\boldsymbol{\alpha}}}{h(\boldsymbol{x})} \right\rangle_{P_t}
\end{equation}
for a multi-index $\boldsymbol{\alpha}$,  and we say each moment has order $|\boldsymbol{\alpha}|$.
Let us define the moment vector $\boldsymbol{y}_{:d}(t)$ as the set of rational moments with order less than or equal to $d$. For a given order $d$ and number of species $S$, the length of the moment vector $\boldsymbol{y}_{:d}(t)$ is $N_d \coloneqq \binom{S+d}{S}$.
The raw and rational moments are related by the linear identity
\begin{align}
 \label{eq:momentrelation}
 \mu^{\boldsymbol{\alpha}}(t)  = {\boldsymbol{\bar{h}}_{\boldsymbol{\alpha}}}^\top \boldsymbol{y}_{:d}(t),
\end{align}
for $d\geq d_h+|\boldsymbol{\alpha}|$, 
indicating that they provide the equivalent information, and $\boldsymbol{\bar{h}}_{\boldsymbol{\alpha}}\in\mathbb{R}^{N_{d}}$ is defined as
\begin{equation}
    \boldsymbol{\bar{h}}_{\boldsymbol{\alpha}}(\boldsymbol{x}) \coloneqq \left\{ \begin{array}{ll}
h_{\boldsymbol{x}-\boldsymbol{\alpha}}, & \text{if } \boldsymbol{x} \geq \boldsymbol{\alpha} \text{ and } |\boldsymbol{x}-\boldsymbol{\alpha}| \leq d_h,\\
0, & \text{otherwise.}
\end{array}\right.
\end{equation}

Denoting the vector of rate constants as
\begin{align}
 \boldsymbol{k} = (k_1,k_2, \ldots, k_R)^\top,
\end{align}
the moment vectors satisfy the \textit{transient moment equations}:
\begin{equation} \label{eqn:mmteqn}
    \frac{\mathrm{d}\mu^{\boldsymbol{\alpha}}}{\mathrm{d}t} = \left\langle Q \boldsymbol{x}^{\boldsymbol{\alpha}} \right\rangle_{P_t} = \boldsymbol{k}^\top A^d_{\boldsymbol{\alpha}} \,  \boldsymbol{y}_{:d}(t),
\end{equation}
for any moment vector with $d \geq |\boldsymbol{\alpha}| +d_b-1$. The coefficient matrix $A^d_{\boldsymbol{\alpha}} \in \mathbb{R}^{R\times N_d}$ can be written in closed-form:
\begin{equation} \label{eqn:A_def}
    A^d_{\boldsymbol{\alpha}}(r,\boldsymbol{l}) \coloneqq \sum_{\substack{\boldsymbol{\gamma} : \ \boldsymbol{l}-d_b  \leq |\boldsymbol{\gamma}| \leq |\boldsymbol{\alpha}|-1} } \binom{\boldsymbol{\alpha}}{\boldsymbol{\gamma}} b_{r,\boldsymbol{l}-\boldsymbol{\gamma}}  \boldsymbol{v}_r^{\boldsymbol{\alpha}-\boldsymbol{\gamma}},
\end{equation}
if $|\boldsymbol{l}| \leq |\boldsymbol{\alpha}|+d_b-1$ and zero otherwise, which depends on the numerator $b_{r}(\boldsymbol{x})$ of the propensity functions and the stoichiometric matrix $\boldsymbol{v}_r$ but not on the reaction parameter $k_r$ (see Appendix~\ref{Sec:mmteqnder} for a derivation). The constant $|\boldsymbol{\alpha}|+d_b-1$ is the order of the highest order moment with a non-zero coefficient involved in the moment equation, which we refer to as the order of the moment equation.

\subsection{Stationary and Generalised Transient Moment Equations}\label{sec:mmteqns}

Here, we seek to construct algebraic constraints between the moments of the reaction network and its parameters. The simplest case is obtained in steady state where the vector of rational moments  $\boldsymbol{y}_{:d}=\lim_{t\to\infty}\boldsymbol{y}_{:d}(t)$ satisfies the \textit{stationary moment equations}:
\begin{equation} 
    \boldsymbol{k}^\top A^d_{\boldsymbol{\alpha}} \,\boldsymbol{y}_{:d} = 0,
\end{equation}
for any moment vector with $d \geq |\boldsymbol{\alpha}| +d_b-1$, which are obtained by setting the time-derivative in \eqref{eqn:mmteqn} to zero. Different algebraic relations hold in the time-dependent case. To this end, we define the \textit{generalised rational moments} as a summary statistic obtained over the time interval $[0,T)$:
\begin{equation} \label{eqn:genrammt}
    \bar{y}^{\boldsymbol{\alpha}}(\rho) \coloneqq \int_{0}^{T} e^{\rho(T-t)} y^{\boldsymbol{\alpha}}(t) dt, 
\end{equation}
where $\rho\in\mathbb{R}$ is a constant. One can similarly define $\boldsymbol{\bar{y}}_{:d}(\rho) $ corresponding to $\boldsymbol{y}_{:d}(t)$ to denote the set of generalised moments up to order $d$. The \textit{generalised transient moment equation} satisfied by these moments are
\begin{align}
    \mu^{\boldsymbol{\alpha}}(T) - e^{\rho T}\mu^{\boldsymbol{\alpha}}(0) + \rho\bar{\mu}^{\boldsymbol{\alpha}}(\rho) = \boldsymbol{k}^\top A^d_{\boldsymbol{\alpha}} \,\boldsymbol{\bar{y}}_{:d}(\rho)
\end{align}
for $|\boldsymbol{\alpha}|\leq d-d_b+1$, which are obtained by multiplying \eqref{eqn:mmteqn} with $e^{\rho(T-t)}$ and using partial integration\citep{dowdy2018dynamic}.  Denoting $\boldsymbol{y}_{:d}(t)$ as $\boldsymbol{y}(t)$ for $t=0,T$ and $\boldsymbol{\bar{y}}_{:d}(\rho)$ as $\boldsymbol{\bar{y}}(\rho)$, and using \eqref{eq:momentrelation}, the generalised transient moment equation can then be written as
\begin{equation}\label{eqn:transmmteqn}
{\boldsymbol{\bar{h}}_{\boldsymbol{\alpha}}}^\top (\boldsymbol{y}(T) - e^{\rho T}\boldsymbol{y}(0) + \rho\boldsymbol{\bar{y}}(\rho)) = \boldsymbol{k}^\top A^{d}_{\boldsymbol{\alpha}} \,
 \boldsymbol{\bar{y}}(\rho) 
\end{equation}
for $|\boldsymbol{\alpha}|\leq d-d_h-d_b+1$. In the following, we will demonstrate how these relations can be used to infer parameters compatible with different moment data.

\section{Theoretical results: Bounds on Rate Parameters via Optimisation}
\label{sec:Method}
In this section, we introduce our approach to defining sets of parameters constrained by the moment equations and moment matrices, and we use outer approximations of these sets to form convex optimisation problems. For steady-state data, we consider two cases of data availability: the \textit{complete data case} (Sec.~\ref{sec:completedata}), in which joint-moments of all species $X_1,\ldots, X_S$ are measured in the experiment, and the \textit{partial data case} (Sec.~\ref{sec:partialdata}), in which moments of only a subset of species are observed. Finally, in Sec.~\ref{sec:transsets}, we extend our approach so that it can be applied when transient data are observed.
 
\subsection{Complete Data Case at Steady State} \label{sec:completedata}

If the true moments are known, one can solve a sufficient number of moment equations directly in \eqref{eqn:mmteqn} for the rate parameters (assuming that the parameters are uniquely identifiable). 
In practice, the true moments are unknown, and we have to use estimates instead. If we were to substitute the true moments with the estimators $\hat{\boldsymbol{y}}$, the moment equations could easily be violated due to sampling errors. We therefore consider moment intervals $(\hat{\boldsymbol{y}}_l,\hat{\boldsymbol{y}}_u)$ that are assumed to contain the true moments and reflect their statistical uncertainty. Such intervals can be obtained by bootstrapping the data, for example, but our approach is not limited to this method (see discussions in Sec \ref{Sec:discussion}).

We can then formulate constrained sets of rate parameters and the true moments compatible with this moment interval data. To this end, we consider a subspace $\mathcal{K} \subseteq {\mathbb{R}_+^R}$ that summarises our prior knowledge about the possible values the rate parameters can assume. The subspace is assumed to have the form
\begin{equation}
    \mathcal{K} \coloneqq \{\boldsymbol{k}\in\mathbb{R}_+^R \ | \ C\boldsymbol{k}=\boldsymbol{c}\}
\end{equation}
for some vector $\boldsymbol{c}$ and matrix $C$. For example, only a subset of parameters can be identifiable in stationary conditions because of the overall timescale that multiplies all rate constants, and hence, we need prior knowledge of some parameters to infer the rest. This means that at least one row of $C$ has a non-zero diagonal entry. Also, some parameters could be constrained to biologically plausible values. This subspace can be more generally defined with inequalities, but we do not consider that case in this paper.

Next, we consider moment equations up to order $d$, namely $|\boldsymbol{\alpha}|+d_b-1\leq d$ with the moment vector $\boldsymbol{y}_{:d}$ (we use $\boldsymbol{y}$ from now onwards for notation simplicity, and the exact moments are denoted as $\boldsymbol{y}^*$). More specifically, we define a constrained set of the following form:
\begin{equation} \label{set:full_y}
 \xi^d = \left\{ {\setstretch{1.314} \begin{array}{l}\boldsymbol{k} \in \mathcal{K}\\ \boldsymbol{y} \in {\mathbb{R}_+^{N_d}} \end{array}}   {\setstretch{1.314} \begin{array}{|ll}
\hat{\boldsymbol{y}}_l \leq \boldsymbol{y} \leq \hat{\boldsymbol{y}}_u
\\
\boldsymbol{k}^\top A^d_{\boldsymbol{\alpha}} \boldsymbol{y} = 0, & |\boldsymbol{\alpha}| \leq d- d_b + 1
\\
 \boldsymbol{y}^\top \boldsymbol{h}_{:N_d} = 1 \\
M_d^s(\boldsymbol{y}) \succeq 0, & s = 0, \ldots, S
\end{array}} \right \}.
\end{equation}
The first inequality is the moment interval data containing the true moment. Note that although we consider bounds on rational moments here,  bounds on raw moments can equivalently be used via \eqref{eq:momentrelation}. The second row is the moment equations satisfied by the true moments and parameters for any $d\geq d_b-1$. The third constraint comes from the fact that $\sum_{\boldsymbol{x} \in \mathcal{X}} \pi(\boldsymbol{x}) = 1$ and thus the rational moments satisfy the condition
\begin{equation} \label{eqn:ys1}
	\boldsymbol{y}_{:d}^\top \boldsymbol{h}_{:N_d} = \sum_{|\boldsymbol{\alpha}| \leq d_h} h_{\boldsymbol{\alpha}} \left\langle \frac{\boldsymbol{x}^{\boldsymbol{\alpha}}}{h(\boldsymbol{x})} \right\rangle_\pi = \left\langle \frac{h(\boldsymbol{x})}{h(\boldsymbol{x})} \right\rangle_\pi = 1
\end{equation}
assuming $N_d \geq d_h$ and $\boldsymbol{h}_{:N_d} \coloneqq (\boldsymbol{h}^\top,0,\ldots,0)^\top$ is a zero-padded version of $\boldsymbol{h}$, the coefficients in $h(\boldsymbol{x})$. In the case of polynomial moments ($h \equiv 1$), this condition becomes $y_0 = 1$. The final constraint arises from the fact that $\boldsymbol{y}$ are moments of a measure over the positive quadrants, with positive semidefinite inequalities associated with the moment matrices\citep{bertsimas2000moment,lasserre2008semidefinite} defined as
\begin{align}\label{eqn:mmtmat}
	&{[M_d^0 (\boldsymbol{y})]_{\boldsymbol{\alpha\beta}} = y_{\boldsymbol{\alpha}+\boldsymbol{\beta}},} 
	\ \ \ \forall\boldsymbol{\alpha},\boldsymbol{\beta} \ \ \textit{s.t.} \ \ |\boldsymbol{\alpha}|,|\boldsymbol{\beta}|\leq \left \lfloor \frac{d}{2} \right \rfloor, \nonumber \\
	&{[M_d^{s} (\boldsymbol{y})]_{\boldsymbol{\alpha\beta}} = y_{\boldsymbol{\alpha}+\boldsymbol{\beta}+\boldsymbol{e}_s},} 
	\ \ \forall\boldsymbol{\alpha},\boldsymbol{\beta} \ \ \textit{s.t.} \ \ |\boldsymbol{\alpha}|,|\boldsymbol{\beta}|\leq \left\lfloor \frac{d-1}{2} \right \rfloor, 
\end{align}
    for $s = 1,\ldots,S$, where $\boldsymbol{e}_i$ is a unit vector whose $i$-th entry is $1$. Here $ M \succeq 0$ denotes the condition where a $n\times n$ symmetric real matrix $M$ is said to be \textit{positive semidefinite}, i.e., $\boldsymbol{v}^\top M \boldsymbol{v} \geq 0$  holds for all $ \boldsymbol{v} \in \mathbb{R}^n$.

Note that if the true moments lie within the moment intervals, the true parameters and the corresponding true moments underlying the data automatically satisfy all constraints; hence the true parameters (and the true moments) are guaranteed to be contained in this set. Thus, if we could optimise each parameter in $\boldsymbol{k}$ over this set, we would obtain theoretically guaranteed lower and upper bounds on the parameters. However,  optimisation over this set is a non-convex, quadratically constrained quadratic programming problem with additional positive semidefinite constraints, and no known solver globally solves this problem. Therefore, we consider a convex outer approximation of this set, over which one can optimise efficiently. The approximation produces potentially looser bounds that still include the true parameters.

To do so, we define the rank-one matrix 
\begin{equation} \label{eqn:zmat}
	Z \coloneqq \boldsymbol{y} \boldsymbol{k}^\top = 
	(\boldsymbol{z}_1 \ \cdots \ \boldsymbol{z}_R)\in \mathbb{R}_+^{N_d \times  R},
\end{equation}
whose columns are moments scaled by a positive rate parameter $\boldsymbol{z}_j \coloneqq k_j\boldsymbol{y} \in {\mathbb{R}_+^{N_d}}$. Using the matrix $Z$, the moment equations in \eqref{eqn:mmteqn} can be rewritten as
\begin{eqnarray} \label{eqn:trace}
    \boldsymbol{k}^\top A^d_{\boldsymbol{\alpha}} \boldsymbol{y} \ && \nonumber = \text{trace}(\boldsymbol{k}^\top A^d_{\boldsymbol{\alpha}} \boldsymbol{y})=\text{trace}( A^d_{\boldsymbol{\alpha}} \boldsymbol{y} \boldsymbol{k}^\top) \\ && =\text{trace}( A^d_{\boldsymbol{\alpha}} Z) = 0
\end{eqnarray}
and the set $\xi^d$ in \eqref{set:full_y} then becomes
\begin{equation} \label{set:full_z1}
	\Tilde{\xi}^d = \left\{{\setstretch{1.314} \begin{array}{l}\boldsymbol{k} \in \mathcal{K} \\ \boldsymbol{y} \in {\mathbb{R}_+^{N_d}} \\ Z \in \\ \ \ \mathbb{R}_+^{N_d\times R}\end{array}} {\setstretch{1.314} \begin{array}{|ll}
			k_j \hat{\boldsymbol{y}}_l \leq \boldsymbol{z}_j \leq k_j \hat{\boldsymbol{y}}_u, & j = 1,\ldots, R \\
			\text{trace}(A^d_{\boldsymbol{\alpha}} Z) = 0, & |{\boldsymbol{\alpha}}| \leq d - d_b + 1 \\
			\boldsymbol{z}_j^\top\boldsymbol{h}_{:N_d} = k_j, & j = 1,\ldots, R\\
			M_d^s(\boldsymbol{y}) \succeq 0, & s = 0, \ldots, S \\
		  ZC^\top = \boldsymbol{y}\boldsymbol{c}^\top\\
            \text{rank}(Z)=1
	\end{array}}\right \}
\end{equation}
where the condition $ZC^\top = \boldsymbol{y}\boldsymbol{c}^\top$ follows from $C\boldsymbol{k}=\boldsymbol{c}$.

So far, this set is equivalent (in terms of $\boldsymbol{k}$ and $\boldsymbol{y}$) to the quadratically constrained problem, and all but the last constraints are linear in the $Z$ and $k$ variables. This suggests that by relaxing the rank-one condition, we can obtain outer approximations of $\Tilde{\xi}^d$ as the following SDP,
\begin{equation} \label{set:full_z}
	\zeta^d  = \left\{{\setstretch{1.314} \begin{array}{l}\boldsymbol{k} \in \mathcal{K}\\ \boldsymbol{y} \in {\mathbb{R}_+^{N_d}} \\ Z \in \\ \ \ \mathbb{R}_+^{N_d\times R} \end{array}} {\setstretch{1.314} \begin{array}{|ll}
		k_j \hat{\boldsymbol{y}}_l \leq \boldsymbol{z}_j \leq k_j \hat{\boldsymbol{y}}_u, & j = 1,\ldots, R\\
			\text{trace}(A^d_{\boldsymbol{\alpha}} Z) = 0, & |{\boldsymbol{\alpha}}| \leq d - d_b + 1\\			\boldsymbol{z}_j^\top\boldsymbol{h}_{:N_d} = k_j, & j = 1,\ldots, R\\
			M_d^s(\boldsymbol{y}) \succeq 0, & s = 0, \ldots, S\\ 
   ZC^\top = \boldsymbol{y}\boldsymbol{c}^\top\\
	\end{array}}\right \}.
\end{equation}

Therefore, the sets~\eqref{set:full_y}, \eqref{set:full_z1}~and~\eqref{set:full_z} form a series of outer approximations with the corresponding bounds:
\begin{align}
    \zeta^d  \supseteq & \Tilde{\xi}^d \supseteq \xi^d, \\
	\label{eqn:filtration_bounds_first}
	\min_{\zeta^d}\boldsymbol{k} \leq \min_{\Tilde{\xi}^d}\boldsymbol{k} \leq \min_{\xi^d}\boldsymbol{k} \leq & \boldsymbol{k}^* \leq \max_{\xi^d}\boldsymbol{k} \leq \max_{\Tilde{\xi}^d}\boldsymbol{k} \leq \max_{\zeta^d}\boldsymbol{k},
\end{align}
where the optimisations are component-wise and so is the comparison between vectors in~\eqref{eqn:filtration_bounds_first}. Hence, one can obtain bounds on the parameters by optimising over $\zeta^d$ with a suitable SDP solver. In Appendix \ref{sec:extraset}, we discuss variations of these schemes with additional moment matrices constraints on $Z$ that tighten the bounds at the expense of computational cost.

\subsection{Partial Data Case at Steady State} \label{sec:partialdata}

A straightforward generalisation relevant in practice is to assume that not all species are observed so that only partial information and intervals for moments associated with the observed species are available.
We split the moment vector $\boldsymbol{y}$ into the \textit{observed moments} ${\boldsymbol{y}}^{obs}$ and the \textit{unobserved moments} $\boldsymbol{y}^{unobs}$, which involves the unobserved species
\begin{equation} \label{eqn:ysplit}
    \boldsymbol{y} \equiv \left(\begin{array}{l}
         \boldsymbol{y}^{obs}  \\
         \boldsymbol{y}^{unobs}
    \end{array} \right),
\end{equation}
and we only have intervals on the observed ones
\begin{equation}
	\hat{\boldsymbol{y}}^{obs}_l \leq {\boldsymbol{y}}^{obs} \leq \hat{\boldsymbol{y}}^{obs}_u.
\end{equation} 
Similarly, we can denote the variables $\boldsymbol{z}_j$ in \eqref{eqn:zmat} as
\begin{equation} \label{eqn:zsplit}
    \boldsymbol{z}_j = \left(\begin{array}{l}
         \boldsymbol{z}_j^{obs}  \\
         \boldsymbol{z}_j^{unobs}
    \end{array}\right) \coloneqq \left(\begin{array}{l}
         k_j\boldsymbol{y}^{obs}  \\
         k_j\boldsymbol{y}^{unobs}
    \end{array} \right), \quad j= 1,\ldots, R.
\end{equation}

A constrained set analogous to $\zeta^d$ in \eqref{set:full_z}, but with partial moment observations, can then be defined as
\begin{widetext}
\begin{equation} \label{set:partial_final}
    \psi^d = \left\{{\setstretch{1.314} \begin{array}{l}\boldsymbol{k} \in \mathcal{K}\\ \boldsymbol{y} \in {\mathbb{R}_+^{N_d}} \\ Z \in \mathbb{R}_+^{N_d\times R} \end{array}} {\setstretch{1.314} \begin{array}{|ll}
    k_j \hat{\boldsymbol{y}}^{obs}_l \leq \boldsymbol{z}^{obs}_j \leq k_j \hat{\boldsymbol{y}}^{obs}_u, & j = 1,\ldots, R\\
\text{trace}(A^d_{\boldsymbol{\alpha}} Z) = 0, & |{\boldsymbol{\alpha}}| \leq d - d_b + 1\\			\boldsymbol{z}_j^\top\boldsymbol{h}_{:N_d} = k_j, & j = 1,\ldots, R\\
			M_d^s(\boldsymbol{y}) \succeq 0, & s = 0, \ldots, S\\ 
   ZC^\top = \boldsymbol{y}\boldsymbol{c}^\top\\
	\end{array}}\right \}.
\end{equation}
\end{widetext}
Clearly, $\zeta^d \subseteq \psi^d$ and hence the bounds obtained from partial moment data are looser than those from full moment data 
since $\zeta^d$ involves more constraints and moment data than $\psi^d$. 

Similar to the bounds on rate parameters, we can obtain bounds on any unobserved moment by optimising over $\psi^d$ with respect to such moment. Following the reasoning for parameters, these lower and upper bounds also satisfy
\begin{equation} \label{eqn:filtration_mmt_bounds}
\min_{\psi^d}y^{\boldsymbol{\alpha}} \leq y^{\boldsymbol{\alpha}*} \leq \max_{\psi^d}y^{\boldsymbol{\alpha}},
\end{equation}
where $y^{\boldsymbol{\alpha}}$ denotes one of the unobserved moments and $y^{\boldsymbol{\alpha}*}$ denotes its true value.

It can be noticed that the created variables in $Z$ are only linked with $\boldsymbol{y}$ through the last equality in $\psi^d$, which depends on the prior constraints we put on $k$. It is thus clear that these constraints allow us to identify and bound unobserved moments. 

\subsection{Generalisation to Time-resolved Data} \label{sec:transsets}

Inference from time-resolved data follows analogously to the steady-state case using the transient moment equations \eqref{eqn:transmmteqn}, which provide relations between the parameters and the generalised moments. Let $\mathcal{P}$ be the selected set of values of $\rho$ from which we compute intervals $(\hat{\boldsymbol{\bar{y}}}_l(\rho),\hat{\boldsymbol{\bar{y}}}_u(\rho))$ of the generalised rational moments in \eqref{eqn:genrammt} from time-dependent data. Additionally, at the start and end points ($t=0,T$) intervals of the rational moments $(\hat{\boldsymbol{y}}_l(t),\hat{\boldsymbol{y}}_u(t))$ are used depending on whether these time points are known or observed.

We create sets of auxiliary variables corresponding to the cross-terms:
\begin{equation} \label{eqn:zmattrans}
	\bar{Z}(\rho) \coloneqq \boldsymbol{\bar{y}}(\rho) (k_0 \ \boldsymbol{k}^\top ) = (\boldsymbol{\bar{z}}_0(\rho) \ \cdots \ \boldsymbol{\bar{z}}_R(\rho))\in \mathbb{R}_+^{N_{d} \times  (R+1)},
\end{equation}
where $\boldsymbol{\bar{z}}_j(\rho) \coloneqq k_j\boldsymbol{\bar{y}}(\rho) \in {\mathbb{R}_+^{N_{d}}}$ for $j=0,\ldots,R$ with $k_0\coloneqq1$ s$\boldsymbol{\bar{z}}_0(\rho)$ is just the generalised moment. Using them, we can express the right-hand side of the transient moment equations \eqref{eqn:transmmteqn} as
$    \boldsymbol{k}^\top A^{d}_{\boldsymbol{\alpha}} \boldsymbol{\bar{y}}(\rho) = \text{trace}(A^{d}_{\boldsymbol{\alpha}} \bar{Z}_{1:R}(\rho))$
following the same steps as in \eqref{eqn:trace} where $\bar{Z}_{1:R}(\rho)$ denotes matrix $\bar{Z}(\rho)$ without the first column.

A suitable outer approximation can then be written as the following SDP:
\begin{widetext}
\begin{equation} \label{set:trans_full_y}
  \tau^d_\mathcal{P} = \left\{ {\setstretch{1.314} \begin{array}{l}\boldsymbol{k} \in \mathcal{K}\\ \boldsymbol{y}(0) \in {\mathbb{R}_+^{N_{d}}}\\
  \boldsymbol{y}(T) \in {\mathbb{R}_+^{N_{d}}}\\
  \text{for } \rho\in\mathcal{P}: \\ \quad \bar{Z}(\rho)\in \mathbb{R}_+^{N_{d} \times  (R+1)}\end{array}}   {\setstretch{1.314} \begin{array}{|ll}
  \text{for }t=0,T:\\
\quad \hat{\boldsymbol{y}}_l(t) \leq \boldsymbol{y}(t) \leq \hat{\boldsymbol{y}}_u(t), & \text{if available}
\\
\quad \boldsymbol{y}^\top(t) \boldsymbol{h}_{:N_{d}} = 1 \\
\quad M_{d}^s(\boldsymbol{y}(t)) \succeq 0, & s = 0, \ldots, S\\
\text{for }\rho \in \mathcal{P}:\\
\quad k_j\hat{\boldsymbol{\bar{y}}}_l(\rho) \leq \boldsymbol{\bar{z}}_j(\rho) \leq k_j\hat{\boldsymbol{\bar{y}}}_u(\rho), & j=0,\dots,R\\
\quad {{\boldsymbol{\bar{h}}}_{\boldsymbol{\alpha}}^\top} (\boldsymbol{y}(T) - e^{\rho T}\boldsymbol{y}(0) + \rho \boldsymbol{\bar{z}}_0(\rho) )= \text{trace}(A^d_{\boldsymbol{\alpha}} \bar{Z}_{1:R}(\rho)), & |\boldsymbol{\alpha}| \leq d -d_h - d_b + 1\\ \quad 
 \boldsymbol{\bar{z}}_j(\rho)^\top \boldsymbol{h}_{:N_{d}} = k_j\frac{e^{\rho T}-1}{\rho} & j=0,\ldots,R\\ \quad M_{d}^s(\boldsymbol{\bar{z}}_0(\rho)) \succeq 0, & s = 0, \ldots, S\\
 \quad 
 \bar{Z}_{1:R}(\rho) C^\top = \boldsymbol{\bar{z}}_0(\rho)\boldsymbol{c}
\end{array}} \right \}
\end{equation}
\end{widetext}
The first inequality represents rational moment interval data at the start and end points, and the second and third lines represent moment and  SDP constraints that they satisfy. The following lines constrain the generalised rational moments on a set of test points that satisfy the transient moment equations. The constraints on $\boldsymbol{\bar{z}}_j(\rho)^\top \boldsymbol{h}_{:N_{d}}$ follow from the corresponding constraints $\boldsymbol{y}^\top(t) \boldsymbol{h}_{:N_{d}} = 1$ and linearity of integration. Note that, if $\rho=0$, the limit needs to be taken, namely $\boldsymbol{\bar{z}}_j(0)^\top \boldsymbol{h}_{:N_{d}} = k_jT$. Similar conditions for the generalised moments, including the SDP inequalities, were obtained in Ref.~\cite{dowdy2018dynamic}.

\section{Numerical results: Application to Biochemical Reaction Networks} \label{sec:Results}
In this section, we apply our approach to biochemical models using synthetic data (complete or partially observed). In the complete data case, we first use the Schl\"ogl model as an introductory example and study the dependence of parameter bounds on the moment order and sample size. We show how to form a constrained set and then compare the bounds on rate parameters obtained from intervals of moments with those based on the true moments. Secondly, we consider the genetic toggle switch with complete data to illustrate that our method can integrate datasets for inferring common parameters. We then consider synthetic data with unobserved species obtained from a model of post-transcriptional gene regulation and demonstrate that datasets with different unobserved species can be combined to tighten parameter bounds and estimate the moments of the unobserved species. Finally, we consider a simple birth-death model with transient data observed. We show that both parameters can be bounded simultaneously with time-course data, and the quality of the bounds depends on the choice of summary statistics and the sample size.

In all steady-state examples, the assumed prior knowledge of the parameters is that at least one parameter is known, i.e., the matrix $C$ is a diagonal matrix with at least one non-zero term and the corresponding value in $c$ is the true parameter value. The datasets used in this section are obtained by collecting samples at the same time point after the burn-in period of $n$ trajectories of Gillespie’s stochastic simulation algorithm \citep{gillespie1977exact}, and assumed to be at steady state. We then use 2000 bootstrap samples of these data points to compute 95\% bootstrap confidence intervals of each moment. In the transient setting, data simulation is described explicitly in Sec. \ref{sec:BDtrans}. We then consider the constrained sets by setting up the semidefinite programs using the package YALMIP \citep{Lofberg2004} and carrying out the optimisation with the solver Mosek \citep{mosek} and default parameters.

\subsection{Schl\"ogl Model with Full Steady-State Observations} \label{example:sch}
Consider the following reaction system \citep{schlogl1972chemical,vellela2009stochastic}:
\begin{equation} \label{model:schlogl}
    2X \xrightleftharpoons[a_2]{a_1} 3X, \quad \varnothing \xrightleftharpoons[a_4]{a_3} X
\end{equation}
with mass-action propensity functions
\begin{eqnarray} \label{eqn:sch_rates}
    &&a_1(x) = k_1x(x-1),\ a_2(x) = k_2x(x-1)(x-2),\nonumber\\ 
    &&a_3(x) = k_3,\  a_4(x) = k_4x.
\end{eqnarray}
In this example, $R = 4$ and all propensities have a common denominator $h\equiv1$, and thus, the rational moments are equivalent to the raw moments. For this model, the highest order of the propensities is $d_b = 3$.
Throughout this subsection, the true value of $\boldsymbol{k}=(k_1, k_2, k_3,k_4)^\top$ is  $(2,3,1,4)^\top$.

To introduce our method with an illustrative example, we first show how the set $\zeta^d$ defined in \eqref{set:full_z} is formed in detail for $d=4$.  In this case, the multi-index is one-dimensional so we denote it with $\alpha$. The moment equations have $\alpha \leq d - d_b + 1 = 2$, and following \eqref{eqn:A_def}, we have the coefficient matrices
\begin{equation} \label{eqn:sch_mmteqn_mat}
	A^{4}_1 = \begin{bmatrix}
		0 & -1 & 1 & 0 & 0 \\
		0 & -2 & 3 & -1 & 0  \\
		1 & 0 & 0 & 0 & 0 \\
		0 & -1 & 0 & 0 & 0
	\end{bmatrix}, A^{4}_2 = \begin{bmatrix}
		0 & -1 & -1 & 2 & 0 \\
		0 & 2 & -7 & 7 & -2  \\
		1 & 2 & 0 & 0 & 0 \\
		0 & 1 & -2 & 0 & 0
	\end{bmatrix},
\end{equation}
which lead to the moment equations as follows
\begin{align} 
\label{eqn:sch_mmteqn1}
\boldsymbol{k}^\top A^4_1 \boldsymbol{y} &=
    k_3y_0 - (k_1+2k_2+k_4)y_1  \nonumber  \\ 
    & + (k_1+3k_2)y_2 - k_2y_3 = 0, \\
 \label{eqn:sch_mmteqn2}
\boldsymbol{k}^\top A^4_2 \boldsymbol{y} &= 
k_3y_0 + (-k_1+2k_2+2k_3+k_4)y_1  \nonumber \\ 
& - (k_1+7k_2+2k_4)y_2 
 + (2k_1+7k_2)y_3 - 2k_2y_4 = 0,
\end{align}
where $y_0 = \langle x^0 \rangle_\pi = 1, y_1 = \langle x^1 \rangle_\pi$ and so on, and the moment equation with $\alpha = 0$ is trivially satisfied. Here, the number of moments is $N_4 = \binom{1+4}{1} = 5$. 

As seen from the above equations, the model does not have moment closure and cannot be solved in closed form. Specifically, the lack of moment closure means that the equation for the first moment depends on higher order moments of order two and three, and the equations for these higher order moments depend on even moments of higher order, thus generating an infinite hierarchy of coupled equations. 

\begin{figure}[ht!] 
\centering
\includegraphics[width=\linewidth]{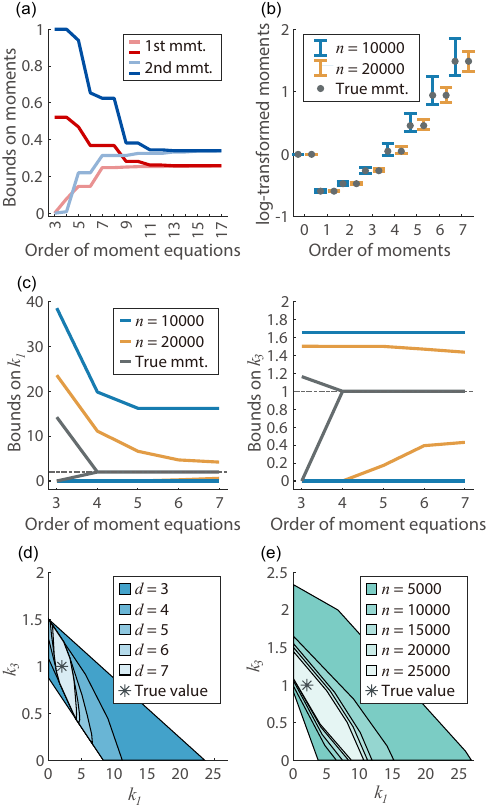}
\caption{\textbf{Parameter inference of the Schl\"ogl model with varying moment order and sample size.} \textbf{(a)} Optimisation-based bounds on $\langle x^1 \rangle_\pi$ and $\langle x^2 \rangle_\pi$ converge to the exact moment (abbr. to mmt.) values. Methods see \citep{kuntz2019bounding}.  \textbf{(b)} Bootstrap intervals of moments, with sample sizes 10000 and 20000, and the true moment values. Values are log-transformed with base 10. \textbf{(c)} Bounds on $k_1$ (left) and $k_3$ (right) obtained from $\zeta^d$ using bootstrap moment intervals with sample sizes 10000 and 20000, and using the exact moments. \textbf{(d)} Feasible regions of $k_1$ and $k_3$ obtained from moment intervals with sample size 20000 for $d = 3,\ldots, 7$. \textbf{(e)} Feasible regions of $k_1$ and $k_3$ obtained from moment intervals with sample sizes varying from 5000 to 25000 and $d = 4$. True parameters are $\boldsymbol{k} = (2,3,1,4)^\top$ and $k_2$ and $k_4$ are assumed to be known in (b) - (e).}
\label{fig:sch_1}
\end{figure}

We first illustrate the forward problem of computing the first two moments using the method described in Ref. \cite{kuntz2019bounding} which provides us with the exact values of the first two moments from convergent bounds on them (Fig.~\ref{fig:sch_1}a). Higher-order moments can then be calculated using the moment equations, assuming the parameters to be known and the first two moments.

In practice, we do not have the exact moments, and we generate a synthetic dataset providing moment intervals via bootstrapping. The log10-transformed bootstrapped moment intervals obtained from datasets of size $10000$ and $20000$ contain the true moments (Fig.~\ref{fig:sch_1}b). As expected, the moment intervals are tighter for low-order moments than for higher-order ones and become tighter when increasing sample size. Note that the bootstrap distributions of the moments are positively skewed for small sample sizes, and thus, the intervals are not symmetric about the true value in the original scale.

Having obtained moment intervals $(\hat{\boldsymbol{y}}_l,\hat{\boldsymbol{y}}_u)$ from the data, we proceed with the inference problem. In the rest of this subsection, we consider $k_2$ and $k_4$ to be known, and we attempt to infer $k_1$ and $k_3$. In this setting, the set $\mathcal{K} = \{\boldsymbol{k}=(k_1,\ldots,k_4)\in\mathbb{R}^4_+ \ | \ k_2=3, k_4=4\}$, namely the matrix $C$ has the second and fourth diagonal term equal to one and all other entries zeros, and the vector $c=(0,3,0,4)^\top$. We then plug this information into the set \eqref{set:full_z}.

In Fig.~\ref{fig:sch_1}c, we show the bounds on both $k_1$ and $k_3$ obtained by optimising $\zeta^d$ with $d = 3,\ldots,7$ using the true moments and bootstrap intervals of moments with the two different sample sizes. We observe that the parameter bounds obtained from the true moments, collapse at order $d=4$. This is because the equation for the first moment \eqref{eqn:sch_mmteqn1} can be satisfied for several choices of $k_1$ and $k_3$, and hence the bounds for $d=3$ reflect the uncertainty in these parameters. When using the first two moment equations, \eqref{eqn:sch_mmteqn1} and \eqref{eqn:sch_mmteqn2}, these can be solved uniquely for $k_1$ and $k_3$ when the first few exact moments are provided. 

Regarding the effect of finite sample size, note that the bounds on $k_1$ tighten as $d$ increases from $3$ to $5$, but beyond order $d=5$ they continue to contract towards the true parameter for sample size $20000$ but not for sample size $10000$. The intuition is that, for smaller sample sizes, the information introduced by including more moment equations after $d = 5$ is not effective since the intervals of the 6th and 7th moments are relatively wide. Similarly, the bounds on $k_3$ do not change much for increasing values of $d$ when the sample size is $10000$. 

To analyse this dependence in more detail, we investigate the feasible parameter regions $\zeta^4$ compatible with the moment interval data up to some order. For a given dataset, the feasible regions depend only on the order of constrained set $d$ and thus on the number of moments used in the inference. Fig.~\ref{fig:sch_1}d shows the feasible regions of $k_1$ and $k_3$ for different orders $d$ with sample size $20000$ and observe that the regions twist as $d$ increases. On the other hand, when $d=4$ is fixed, but the sample size varies, the regions shrink with the sample size, but their shapes are generally similar (Fig.~\ref{fig:sch_1}e). In summary, the tightness of parameter bounds is limited by the orders of moments available for inference and the sample size.

\subsection{Toggle Switch Model with Multiple Full Observations at Steady State}

Next, we demonstrate our approach for the case of parameter inference with multiple perturbation datasets. Perturbations here describe parameter changes due to conditions that represent experimental changes in gene induction levels or CRISPR perturbations in cells. We consider the toggle switch model \cite{gardner2000construction} with two chemical species and rational propensities:
\begin{equation} \label{model_tg}
    \varnothing \xrightarrow{a_1} X_1 \xrightarrow{a_2} \varnothing, \quad \varnothing \xrightarrow{a_3} X_2 \xrightarrow{a_4} \varnothing
\end{equation}
with propensities
\begin{eqnarray}\label{eqn:tg_rates}
    &&a_1(\boldsymbol{x}) = \frac{k_1(1+x_1)}{(1+x_2^3)(1+x_1)}, \nonumber \\ &&a_2(\boldsymbol{x}) = \frac{k_2x_1(1+x_2^3)(1+x_1)}{(1+x_2^3)(1+x_1)}, \nonumber \\
    &&a_3(\boldsymbol{x}) = \frac{k_3(1+x_2^3)}{(1+x_2^3)(1+x_1)}, \nonumber \\
 &&a_4(\boldsymbol{x}) = \frac{k_4x_2(1+x_2^3)(1+x_1)}{(1+x_2^3)(1+x_1)},
\end{eqnarray}
where all rate parameters are positive. Here, $d_b=5$ and the common denominator $h(\boldsymbol{x})$ has the form
\begin{equation}
    h(\boldsymbol{x}) = (1+x_2^3)(1+x_1)
\end{equation}
with $d_h = 4$. The rational moments $\boldsymbol{y}^{\boldsymbol{\alpha}}$ can be defined and calculated correspondingly. 

We use this model to illustrate how to integrate datasets from different numerical experiments. To this end, we consider the toggle switch model \eqref{model_tg} with two parameter settings: $\boldsymbol{k}_1 = (k_{1,1}, k_{1,2},k_{1,3},k_{1,4})$ and $\boldsymbol{k}_2 = (k_{2,1}, k_{2,2},k_{2,3},k_{2,4})$ with moment intervals $(\hat{\boldsymbol{y}}_{1,l}, \hat{\boldsymbol{y}}_{1,u})$ and $(\hat{\boldsymbol{y}}_{2,l}, \hat{\boldsymbol{y}}_{2,u})$, respectively. Using these, we define two constrained sets $\zeta^d_1\coloneqq \zeta^d(\hat{\boldsymbol{y}}_{1,l}, \hat{\boldsymbol{y}}_{1,u},\mathcal{K}_1)$ and $\zeta^d_2\coloneqq \zeta^d(\hat{\boldsymbol{y}}_{2,l}, \hat{\boldsymbol{y}}_{2,u},\mathcal{K}_2)$, where $\mathcal{K}_1 = \{\boldsymbol{k}_1=(k_{1,1},\ldots,k_{1,4})\in\mathbb{R}^4_+ \ | \ k_{1,1}=k_{1,1}^*, k_{1,2}=k_{1,2}^*\}$, and similar for $\mathcal{K}_2$. Here $k_{1,3} = k_{2,3}$ and $k_{1,4} = k_{2,4}$ are shared parameters which we want to infer, and the other parameters are perturbed but assumed to be known for simplicity.  We can then combine the sets as $ \{ \boldsymbol{k}_1 \in \zeta^d_1, \boldsymbol{k}_2 \in \zeta^d_2 |  {k}_{1,3} = {k}_{2,3}, {k}_{1,4} = {k}_{2,4} \} $. 
We can combine more than two datasets by repeatedly applying this method.

\begin{figure}[htb!] 
\centering
\includegraphics{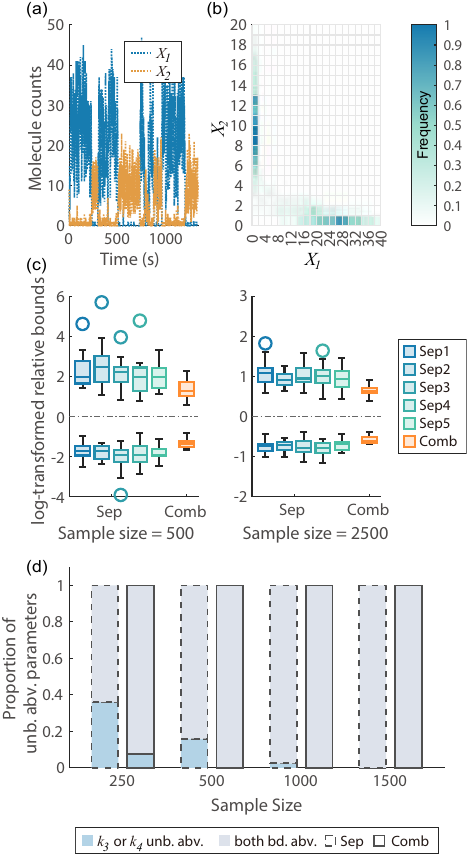}
\caption{\textbf{Data integration for parameter inference of the toggle switch model with rational propensities.} \textbf{(a)} Trajectories of the two species in the system. Parameters are Par1. \textbf{(b)} Histogram of a dataset generated with parameters Par1.  \textbf{(c)} Box plots of natural-log-transformed relative bounds on $k_3$ with different methods and sample sizes (left: $n=500$, right: $n=2500$). Experiments are repeated 20 times. The bounds Sep\,\textit{i} are obtained by using the dataset generated with parameters Par\,\textit{i} and the bounds Comb are obtained by combining the five datasets used for Sep. Unbounded-above (abbr. to unb. abv.) issues are omitted for Sep with a sample size of 500. Parameters used here are Par1 (20, 0.7, 10, 1), Par2 (24, 0.82, 10, 1), Par3 (30, 1.1, 10, 1), Par4 (22, 0.8, 10, 1) and Par5 (28, 0.9, 10, 1). \textbf{(d)} The proportion of experiments resulting in at least one parameter unbounded-above compared with both parameters bounded-above (abbr. to bd. abv.) when inference is performed using each of the five datasets separately or using their combination. The sample size of each dataset increases from 250 to 1500, and the experiments are repeated 20 times for each sample size. The parameter settings are the same as in (c). In panels (c) and (d), the parameters $k_1$ and $k_2$ are assumed to be known and $d=6$.} 
\label{fig:tg_1}
\end{figure}

As an illustration, we generate datasets with five different parameter settings in which the values of $k_{m,3}$ and $k_{m,4}$ are fixed for $m=1,\ldots,5$ and we refer to them as $k_3$ and $k_4$ later on. These are the parameters we intend to infer, and the other two parameters, $k_{m,1}$ and $k_{m,2}$, are assumed to be known but changing across the settings. It can be seen in Fig.~\ref{fig:tg_1}a,b that, with parameter setting Par1 (defined in its caption) as an example, the trajectories show the expected switching behaviour and the stationary distribution of each of the two species is bimodal. The behaviour for the other four parameter settings is omitted here as they are very similar. In particular, five datasets of size 500 and five datasets of size 2500 are generated with each of the five parameter settings for comparison. For a fixed sample size, bounds on the parameters can then be obtained by separately using each dataset (referred to as $\text{Sep1,}\ldots\text{, Sep5}$) or by implementing the method introduced above to combine all five datasets (referred as Comb). 

We repeat the experiment 20 times, namely generating these ten datasets and performing inference with different methods, and produce box plots of the natural-log-transformed relative bounds in Fig.~\ref{fig:tg_1}c. In addition to noting that the bounds consistently contain the true parameter, the medians of the transformed relative upper (\textit{resp.} lower) bounds on $k_3$ obtained by using each dataset with sample size 500 separately concentrate around $2$ (\textit{resp.} $-1.5$) while the median of those obtained by combining the datasets concentrate around $1$ (\textit{resp.} $-1$). Thus the combination of datasets significantly tightens the bounds and decreases their variance across the repetitions. Qualitatively similar results are obtained when the sample size increases to 2500 but all relative bounds are nearly halved on the log-transformed scale regardless of the method used. The outcomes for $k_4$ are very similar and hence are omitted.

We found that bounding some parameters from above can be infeasible for small sample sizes due to loose moment intervals. We can observe from Fig \ref{fig:tg_1} (d) that combining datasets alleviates the potential unbounded-above problems, especially for smaller sample sizes. For instance, when each of the five datasets has only 250 samples, in around 40\% experiments, at least one parameter cannot be bounded from above, whilst this proportion falls below 10\% if combined. Thus, integrating datasets of different parameter settings leads to tighter parameter bounds in all settings.

\subsection{Post-Transcriptional Gene Regulation with Partial Observations at Steady State} \label{sec:post-tran}
We apply our method to networks with unobserved species, which we refer to as the partial data case. As a simplified model of post-transcriptional regulation in cells \citep{statello2021gene,hausser2014identification} with bimolecular reactions, we consider the following 2-species model:
\begin{equation} \label{model_sp2}
    \varnothing \xrightleftharpoons[a_2]{a_1} X_1, \quad \varnothing \xrightleftharpoons[a_4]{a_3} 2X_2, \quad X_1 + X_2 \xrightarrow{a_5} \varnothing,
\end{equation}
with mass-action propensity functions
\begin{eqnarray} \label{eqn:sp2_rates}
    &&a_1(\boldsymbol{x}) = k_1, \ a_2(\boldsymbol{x}) = k_2x_1, \ a_3(\boldsymbol{x}) = k_3,\nonumber \\ &&a_4(\boldsymbol{x}) = k_4x_2(x_2-1), \ a_5(\boldsymbol{x}) = k_5x_1x_2.
\end{eqnarray}  
Given an unobserved species, we use bootstrap samples of the observed data to form intervals of the moments, which only include the observed species. For instance, if $X_1$ is observed but $X_2$ is not, only the moment intervals of $\langle x_1^\alpha x_2^0 \rangle_\pi$ are available.

\begin{figure}[htb!]
\centering
\includegraphics{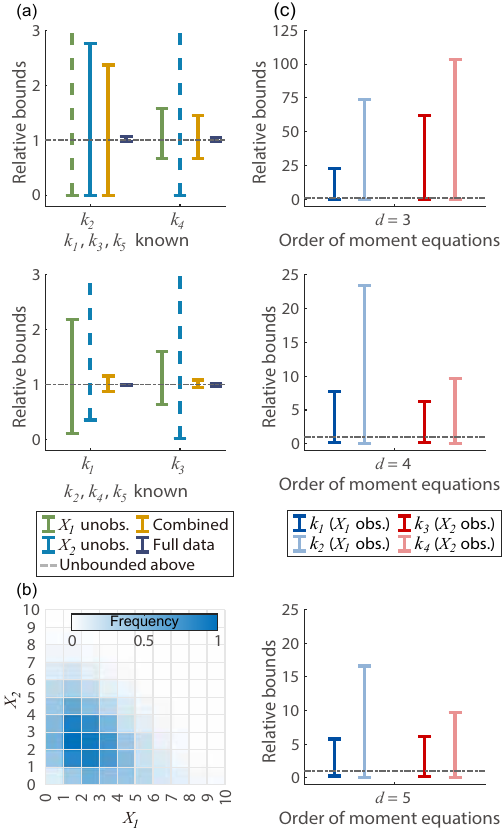}
\caption{\textbf{Parameter bounds of the post-transcriptional regulation model with partial data.} \textbf{(a)} Relative bounds on the unknown parameters when $k_1,k_3,k_5$ are known (left) and when $k_2,k_4,k_5$ are known (right) with $d=4$. In each case, we compare the bounds obtained when $X_1$ is unobserved; when $X_2$ is unobserved; when we have two datasets each with only one species measured; and when we have full data. \textbf{(b)} Histogram of $X_1$ and $X_2$ showing they have similar magnitudes at steady state. \textbf{(c)} Relative bounds on parameters obtained with $d = 3,4,5$. We compare the bounds on the unknown parameters when assuming $k_3,k_4,k_5$ known but $X_1$ observed and when assuming $k_1,k_2,k_5$ known but $X_2$ observed. In all panels, the true parameters are $\boldsymbol{k} = (6,0.8,5,0.5,1)^\top$. The sample size is $n=20000$ in panels (a) and (c).}
\label{fig:sp2_1}
\end{figure}

We consider the test case in which we have two synthetic datasets of the same model and the same parameters, but one only has observations on $X_1$ and the other only measures $X_2$. Inference can be performed with these datasets separately with different unobserved species. These two datasets can be used to provide moment intervals on  $\langle x_1^\alpha x_2^0 \rangle_\pi$ and  $\langle x_1^0 x_2^\alpha \rangle_\pi$ respectively, which can be used to constrain the sets. 

One can try to integrate these two datasets. Since the data are collected independently, one can only obtain intervals on the marginal moments but no information on the cross-moments. By inputting these marginal intervals into the sets, more constraints on the dynamics are introduced, and the bounds are expected to be tighter than when using each dataset separately. 

We then compare the tightness of the resulting bounds with those obtained in the full data case, in which we have both the marginal and cross-moment intervals and those obtained by using each dataset separately (Fig.~\ref{fig:sp2_1}a). We see that the bounds obtained with full data are very tight around the true values, and if datasets are not combined, the bounds are much looser or unbounded above. Combining the two datasets, it can be observed that the parameters can always be bounded above when assuming different known parameters, and the bounds are tighter than using each data separately. This effect is more significant when only the three decay rates are known, where the bounds are almost as tight as in the full data case.

In Fig.~\ref{fig:sp2_1}c, we instead assume that the parameters in the production and decay reactions of the unobserved species, as well as $k_5$, are known and try to infer those of the observed species. Note that we do not consider the combination of datasets in this example. Overall, the bounds tighten as $d$ increases regardless of which species is measured. For all $d$ and observed species, we can see that the bounds on the degradation rates $k_2$ and $k_4$ are tighter than those on the production rates $k_1$ and $k_3$. This is because the degradation rates are multiplied with higher order moments in the moment equations and these moments have wider moment intervals. Comparing the cases in which $d$ is fixed but different species are observed, it can be seen that, for higher orders, the bounds on the degradation rate are tighter when $X_2$ is observed than when $X_1$ is observed. Although various reasons may cause such observations, the most likely one is that $k_2$ is multiplied with higher order moments as the reaction involves two molecules, and the constraints on these moments are less than those on the lower order moments. Therefore, introducing moment intervals on them provides information on the less certain variables in the system and the corresponding bounds are tighter. We choose the parameters to ensure that these observations are not due to varying magnitudes of the stationary moments (Fig.~\ref{fig:sp2_1}b).

\begin{figure}[htb!]
\centering
\includegraphics{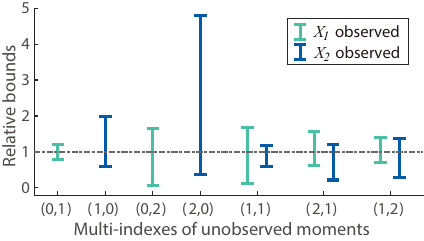}
\caption{\textbf{Inference of moments of the unobserved species in the post-transcriptional regulation model.} Parameters $(k_3,k_4,k_5)$ and $(k_1,k_2,k_5)$ are assumed to be known, respectively, when $X_1$ and $X_2$ are observed. The sample size is $n=20000$, the order $d$ is 3, and the true parameters are the same as in Fig.~\ref{fig:sp2_1}. The relative bounds are obtained by dividing the original bounds by bootstrap intervals of the corresponding moments.}
\label{fig:sp2_2}
\end{figure}

In addition to the inference of parameters, we demonstrate that our approach can also be used to bound the stationary moments of the unobserved species
\begin{align}
	\min_{\psi^d}y^{\boldsymbol{\alpha}} \leq y^{\boldsymbol{\alpha}*} \leq  \max_{\psi^d}y^{\boldsymbol{\alpha}}
\end{align}
where $y^{\boldsymbol{\alpha}}$ denotes one of the unobserved moments and $y^{\boldsymbol{\alpha}*}$ denotes its true value. The bounds on these moments are divided by the mean of bootstrap intervals for comparison (Fig.~\ref{fig:sp2_2}). Although the quality of the bounds decreases with the order of the moment, the bounds on the means or cross moments are relatively tight and reasonably constrain the moments of the latent species. This demonstrates that our approach provides quantified uncertainties for parameters and moment predictions of stochastic reaction networks with latent species.

\subsection{Birth-Death Process with Time-dependent Observations} \label{sec:BDtrans}
\begin{figure}[ht!]
\centering
\includegraphics{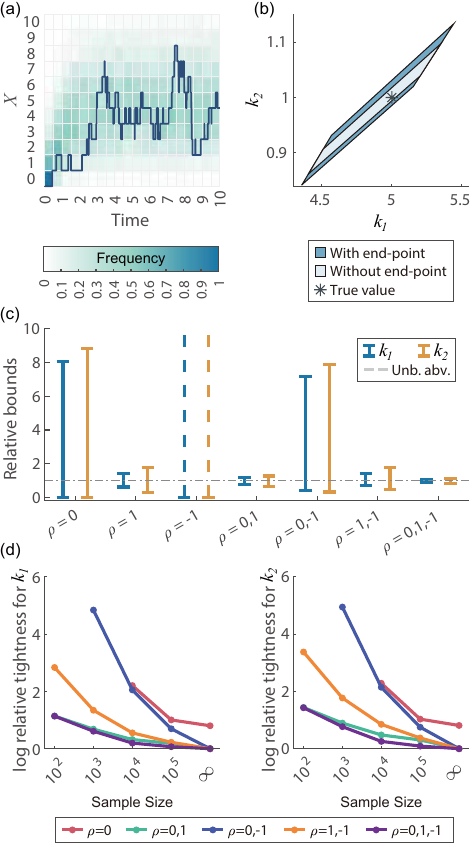}
\caption{\textbf{Parameter inference of a birth-death model with time-dependent data.} \textbf{(a)} Data simulated from the birth-death model \eqref{model_bd} with sampling time uniformly distributed between 0 and 10. These data are used to obtain intervals on the generalised moments. An example trajectory of the dynamic is also plotted. \textbf{(b)} Feasible regions of $k_1$ and $k_2$ with and without 5000 additional observations at the end-point $T$ using three values of $\rho=0,1,-1$. \textbf{(c)} Bounds on both parameters with different choices of $\rho$. Dashed lines indicate that the parameter is unbounded above. \textbf{(d)} The log relative tightness for different sample sizes, which is calculated as $\text{log}((k_u-k_l)/k+1)$ where $k_u$ and $k_l$ are the upper and lower bounds we obtain for a parameter $k$. A missing value of the tightness indicates that the parameter is unbounded above. Infinite sample size ("$\infty$") corresponds to inference performed with exact moments computed from the time-varying Poisson solution. In panels (a) - (c), the sample size of the transient data is 10000. In panels (c) and (d), no data at the end-point are considered. The true parameters are $\boldsymbol{k}=(5,1)^\top$ and $T=10$ in all panels and all optimisation was done with $d=3$ with intervals obtained from $2,000$ bootstrap samples.}
\label{fig:bdtrans}
\end{figure}

To illustrate the use of our method for transient data, we consider a simple birth-death process:
\begin{equation} \label{model_bd}
    \varnothing \xrightleftharpoons[a_2]{a_1} X,
\end{equation}
with mass-action propensity functions
\begin{equation}
    a_1(x) = k_1, \ a_2(x) = k_2x.
\end{equation}  
We generate test data by sampling time courses using the SSA starting from $x=0$. To obtain intervals on the generalised moments, we take observations at sample time points $t\sim\text{Uniform}(0,T)$ (Fig.~\ref{fig:bdtrans}a). We then use bootstrapping to obtain intervals on the generalised moments $\boldsymbol{\bar{y}}(\rho)$ for a selected set of values of $\rho$, and on the moments at the endpoints.

We observe that the feasible regions of $k_1$ and $k_2$ are bounded tightly around the true parameters and show a positive correlation between the two parameters (Fig.~\ref{fig:bdtrans}b). When moment data at the endpoint are used, more constraints are introduced, and the feasible region shrinks compared to generalised moments only. 
To study the influence of the summary statistics on the tightness of the bounds,
we consider different sets of $\rho$ taken as combinations of $\rho = 0,1,-1$  (Fig.~\ref{fig:bdtrans}c). We see that the parameters cannot be bounded from above for $\rho=-1$ while they are bounded for non-negative values (e.g. $\rho = 0$ or $\rho = 1$). Including more values of $\rho$ tightens bounds, and the best bounds are obtained when combining all three values.

In Fig.~\ref{fig:bdtrans}d, we study the convergence of our bounds with sample size.
The bounds approach the true parameters as the sample size is increased, and they converge faster as more $\rho$ values are included. To demonstrate this, we use the fact that the transient solution of the CME is a time-varying Poisson distribution and compute the generalised moments analytically (nominally corresponding to an infinite sample size). Only for $\rho=0$, the bounds do not converge to the true parameters even with the exact values of the generalised moments. Intuitively, negative values of $\rho$ put more weight at the end of the time series; positive values emphasise the transients at the beginning; and zero values average across time. Hence a suitable range of $\rho$ needs to be selected on a case-by-case basis considering the time scale of the dynamics. Our findings highlight that the choice of summary statistics influences identifiability, bound tightness, and convergence.  

\section{Discussion} \label{Sec:discussion}

We proposed a novel inference approach for stochastic reaction networks based on moment constraints and mathematical programming. We used the moment equations and the moment matrices to form a constrained set of the rate parameters and outer approximate this set by introducing variables to replace the non-linear terms in the constraints. Upper and lower bounds on the parameters can then be obtained by optimising over this set, forming a convex semidefinite programming problem with a unique optimum. Our method takes intervals of the stationary moments or the generalised moments as input, and the bounds obtained are guaranteed to contain the true parameters under the assumption that these intervals include the true moments.

In contrast to likelihood-based and simulation-based approaches, our approach avoids the solution of the forward problem, i.e., predicting moments and distributions for given parameters, and thus avoids intractable likelihood computations or computationally expensive simulations associated with these approaches. When the solution is known, our result converges to the interval of the method of moments estimates with bootstrap resampling, but our approach has the advantage that the moment equations do not need to be solved explicitly. Unlike likelihood- and simulation-based approaches, our approach is guaranteed to converge in polynomial time and its computational cost scales only with the number of parameters and the order of the constrained set \citep{traub1982complexity, vandenberghe1996semidefinite}. Moreover, our approach enjoys the advantage that it provides bounds on the parameters and thus provides error guarantees under mild assumptions. 

Several methods to infer parameters using moment equations have been proposed in the literature. For instance, \emph{Backenkohler et al.}\cite{backenkohler2017moment} minimise a weighted sum of squares of the moment equations using point estimates of the moments. This approach provides point estimates of the parameters, and it thus remains unclear how accurate its parameter estimates and predictions are. Other approaches write likelihoods for moments; these are however, tractable only using additional approximations, such as moment closure or the system size expansion, that can introduce uncontrolled errors and have to be validated on a case-by-case basis\cite{zechner2012moment,ruess2015moment,frohlich2016inference}. Even in special cases when the moments are tractable, sampling errors can lead to significant bias or even negative parameter estimates \cite{tang2023modelling}. Our method enjoys the advantage that the moment equations do not need to be solved explicitly, i.e., it avoids the forward problem. Hence, our method will be widely applicable even to complex stochastic reaction networks.

Our approach represents a computationally tractable and efficient convex outer approximation to the original optimisation problem, which falls into the broad and complicated category of non-convex quadratically constrained quadratic programs. We have demonstrated that our approximation generally provides tight and informative bounds in a number of applications. Yet it remains unclear what conditions guarantee
the tightness of our bounds. For example, it is likely that for inference problems with partial observations, products of unknown parameters and unobserved moments cannot be identified using the linearisation involved in our outer approximations. In such cases, the original non-convex inference problem can produce tighter bounds with a different optimisation solver and potentially much larger computational cost. Furthermore, the present method can be extended to include observation errors by coupling true and noise-corrupted moments (see Appendix~\ref{sec:obserr}). Unless the hyperparameters of the noise model are known, estimating the additional noise parameters leads to a non-convex cubically constrained program that can also suffer from poor identifiability under linearisation. Further work will be needed to investigate these issues.

An extension of our method would be to utilise moment intervals other than bootstrap intervals as input data. For example, one could compute concentration inequalities to couple tail probabilities with moment equations. Another possible development would be to consider parameter distributions instead of point estimates of parameters. Such parameter heterogeneity arises, for example, due to extrinsic noise in cells \cite{zechner2012moment,dattani2017stochastic,thomas2021coordination} and can be implemented in our framework by adding extra static species, which are not changed in reactions. Inference of moments of these parameters then proceeds by producing bounds on moments of the unobserved species analogue to the partial data case (Fig.~\ref{fig:sp2_2}). The corresponding forward problem that bounds moments given known parameter distributions has been discussed by \emph{Sakurai and Hori}\citep{sakurai2022interval}, and the same constraints could be used in our inverse problem to infer bounds on moments of the parameter distributions. This approach introduces additional variables and equations for cross-moments of parameters and species, thus adding to the complexity of the optimisation problem.

Here, we employed raw moments for measurements at steady state and generalised moments for time-course inference. In the latter case, generalised moments serve as summary statistics of time courses where the values of the exponent $\rho$ control the weight given to early and late time points. We have shown that these exponents must be chosen carefully and evaluated case-by-case to obtain tight parameter bounds. We also note that the use of generalised moments differs from instantaneous moments taken at discrete time intervals in standard setups \citep{zechner2012moment,ruess2015moment,davidovic2022parameter}. Still, the generalised moments can be interpreted as a time-averaged version of the instantaneous ones when these are sufficiently frequently sampled and thus provide similar information. It remains an open question whether employing other statistics could improve the sufficiency, convergence properties and computational cost of our parameter inference method.

Our method could be further used to explore the design space of synthetic biochemical circuits within the context of the work of \emph{Sakurai and Hori} ~\cite{sakurai2018optimization}. Our approach could be applied to that problem if one uses the desired output moments of a synthetic circuit as moment intervals. Optimising over parameters then provides bounds on the feasible parameter space, thus circumventing the simulation effort needed for the forward problem when designing circuits with defined statistical properties.

In conclusion, the present method offers a versatile and extensible framework for parameter inference and prediction using stochastic reaction networks. Unlike conventional parameter estimation methods, the framework enjoys the advantage of providing quantified error bounds and theoretical guarantees on computational efficiency and convergence. Given the prevalence of stochastic reaction networks across disciplines, we expect that our advances will be transformative across applications that rely fundamentally on speed and accuracy guarantees of model predictions.
\subsection*{CODE AVAILABILITY}
MATLAB code is available at \url{github.com/pthomaslab/SDP-inference}.

\begin{acknowledgments}
We thank Ruth Misener for insightful discussions. ZL is supported through a PhD studentship by the Department of Mathematics at Imperial College London. MB acknowledges support by the EPSRC under grant EP/N014529/1 funding the EPSRC Centre for Mathematics of Precision Healthcare at Imperial.  PT was supported by UKRI through a Future Leaders Fellowship (MR/T018429/1).
\end{acknowledgments}

\appendix
\section{Derivation of the Moment Equations} \label{Sec:mmteqnder}
Assuming that $P_t$ satisfies the chemical master equation, the equation
\begin{equation} \label{eqn:mmteqnA}
    \frac{\mathrm{d}}{\mathrm{d}t} \langle x^{\boldsymbol{\alpha}} \rangle_{P_t}  = \left\langle Q\boldsymbol{x}^{\boldsymbol{\alpha}}  \right\rangle_{P_t} \coloneqq \sum_{\boldsymbol{x}'\in\mathcal{X}} \sum_{\boldsymbol{x}\in\mathcal{X}} q(\boldsymbol{x}',\boldsymbol{x}) \boldsymbol{x}^{\boldsymbol{\alpha}} P_t(\boldsymbol{x}') 
\end{equation}
holds for all $\boldsymbol{\alpha} \in \mathbb{N}^S$. Following Ref. \cite{kuntz2019bounding} one can then write
\begin{eqnarray} 
	\label{eqn:app_momeq}
    h(\boldsymbol{x})Q\boldsymbol{x}^{\boldsymbol{\alpha}} && = \sum_{r=1}^R k_r b_r(\boldsymbol{x})((\boldsymbol{x}+\boldsymbol{v}_r)^{\boldsymbol{\alpha}}-\boldsymbol{x}^{\boldsymbol{\alpha}})\nonumber \\
    && = \sum_{r=1}^R k_r \sum_{|\boldsymbol{\beta}|\leq d_b}b_{r,\boldsymbol{\beta}} \boldsymbol{x}^{\boldsymbol{\beta}} \sum_{|\boldsymbol{\gamma}| \leq |\boldsymbol{\alpha}|-1} \binom{\boldsymbol{\alpha}}{\boldsymbol{\gamma}} \boldsymbol{x}^{\boldsymbol{\gamma}} \boldsymbol{v}_r^{\boldsymbol{\alpha}-\boldsymbol{\gamma}} \nonumber \\
   && = \sum_{r=1}^R k_r \sum_{|\boldsymbol{\beta}|\leq d_b} \sum_{|\boldsymbol{\gamma}| \leq |\boldsymbol{\alpha}|-1} \binom{\boldsymbol{\alpha}}{\boldsymbol{\gamma}} b_{r,\boldsymbol{\beta}}   \boldsymbol{v}_r^{\boldsymbol{\alpha}-\boldsymbol{\gamma}} \boldsymbol{x}^{\boldsymbol{\beta}+\boldsymbol{\gamma}} \nonumber \\
    && = \sum_{r=1}^R k_r \sum_{|\boldsymbol{l}|\leq d} A^d_{\boldsymbol{\alpha}}(r,\boldsymbol{l}) \boldsymbol{x}^{\boldsymbol{l}},
\end{eqnarray}
where the elements in the coefficient matrix $A^d_{\boldsymbol{\alpha}} \in \mathbb{R}^{R\times N_d}$ are defined as in \eqref{eqn:A_def}. Dividing \eqref{eqn:app_momeq} by $h(\boldsymbol{x})$ and taking expectations it follows that 
\[
	\frac{\mathrm{d}}{\mathrm{d}t}\mu^{\boldsymbol{\alpha}} = \left\langle Q\boldsymbol{x}^{\boldsymbol{\alpha}} \right\rangle_{P_t} = \boldsymbol{k}^\top A^d_{\boldsymbol{\alpha}} \boldsymbol{y}_{:d},
\] which are the moment equations \eqref{eqn:mmteqn} in the main text.

\section{Additional Moment Matrices Constraints at Steady State} \label{sec:extraset} 

Following the definition of $Z$ in \eqref{eqn:zmat}, its columns are moments scaled by a positive rate parameter $\boldsymbol{z}_j \coloneqq k_j\boldsymbol{y} \in {\mathbb{R}_+^{N_d}}$
that inherit the positive semidefinite properties
\begin{equation} \label{eqn:extrammtmat}
	M_d^s(\boldsymbol{z}_j) \equiv k_jM_d^s(\boldsymbol{y}) \succeq 0, \quad s = 0, \ldots, S
\end{equation}
for $j=1,\ldots,R$.

Note that, in the set $\Tilde{\xi}^d$ defined in \eqref{set:full_z1}, these extra positive semi-definite constraints are automatically fulfilled because of the rank-one condition and the parameter(s) that we assume to be known. Hence, instead of removing the rank-one condition directly, a semi-definite relaxation will be replacing it with these constraints in \eqref{eqn:extrammtmat} and the extent of this relaxation will depend on how many moment matrices constraints are included.

Considering a permutation $\sigma$ of the index set $\{1,2,\ldots,R\}$, we then obtain the outer approximations
\begin{equation} \label{set:tighter_z}
	\zeta^d_J  = \left\{{\setstretch{1.314} \begin{array}{l}\boldsymbol{k} \in \mathcal{K}\\ \boldsymbol{y} \in {\mathbb{R}_+^{N_d}} \\ Z \in \\ \ \ \mathbb{R}_+^{N_d\times R} \end{array}} {\setstretch{1.314} \begin{array}{|ll}
 
			k_j \hat{\boldsymbol{y}}_l \leq \boldsymbol{z}_j \leq k_j \hat{\boldsymbol{y}}_u, & j = 1,\ldots, R\\
			\text{trace}(A^d_{\boldsymbol{\alpha}} Z) = 0, & |{\boldsymbol{\alpha}}| \leq d - d_b + 1\\
			\boldsymbol{z}_j^\top\boldsymbol{h}_{:N_d} = k_j, & j = 1,\ldots, R\\
			M_d^s(\boldsymbol{y}) \succeq 0, & s = 0, \ldots, S\\
   ZC^\top = \boldsymbol{y}\boldsymbol{c}^\top\\
   M_d^s(\boldsymbol{z}_{\sigma(j)}) \succeq 0, & s = 0, \ldots, S, \\
			 & j=1,\ldots,J
	\end{array}}\right \},
\end{equation}
where the index $J \leq R$ denotes the number of moment matrices constraints included.

These sets form a descending filtration of outer approximations
\begin{align}
	\label{eqn:filtration}
	\xi^d \subseteq \zeta^d_R \subseteq \cdots \subseteq \zeta^d_2  \subseteq \zeta^d_1 \subseteq \zeta^d,
\end{align}
for a given permutation $\sigma$. This property implies the upper and lower bounds on the true parameter
\begin{align}
	\label{eqn:filtration_bounds}
	\min_{\zeta^d}\boldsymbol{k} \leq \min_{\zeta^d_1}\boldsymbol{k} \leq \cdots \leq  \min_{\zeta^d_R} \boldsymbol{k} \leq  \min_{{\xi}^d} \boldsymbol{k}
	 \leq \boldsymbol{k}^*  \notag\\
	 \qquad
	  \leq \max_{{\xi}^d} \boldsymbol{k} \leq 
	 \max_{\zeta^d_R}\boldsymbol{k} \leq \cdots \leq  \max_{\zeta^d_1}\boldsymbol{k} \leq \max_{\zeta^d}\boldsymbol{k}.
\end{align}

The corresponding bounds of $\zeta^d_J$, $\zeta^d$ and $\xi^d$ are equal if the stationary moments are known exactly, i.e. when there is no statistical uncertainty. More generally, if an optimum of $\zeta^d_1$ or $\zeta^d$ is also a feasible point of $\zeta^d_R$, i.e., it satisfies all additional semidefinite constraints, then this is also an optimal point of $\zeta^d_R$. Similarly, if the optimum also satisfies the rank-one constraint, it is also an optimal point of $\Tilde{\xi}^d$ (and $\xi^d$). In addition, it is clear that 
$\zeta^1_J \subseteq \cdots \subseteq \zeta^{d}_J$, which implies that bounds do not loosen
\begin{align}
	\label{eqn:filtration_bounds2}
	\min_{\zeta^{1}_J}\boldsymbol{k} \leq \cdots \leq  \min_{\zeta^d_J} \boldsymbol{k} 
	\leq \boldsymbol{k}^\ast  
	\leq 
	\max_{\zeta^d_J}\boldsymbol{k} \leq \cdots \leq  \max_{\zeta^1_J}\boldsymbol{k}.
\end{align}
with an increasing number of moment intervals available, and similar for other $\xi^d, \Tilde{\xi}^d$ and $\zeta^d$. 

The sets $\zeta^d$ and $\zeta^d_J$ contain the true parameters (assuming moment intervals contain the true moments), as does $\xi^d$. However, they are considerably simpler to optimise since bounds on parameters can be obtained using convex optimisation tools in polynomial time and smaller values of $J$ provide the most computationally efficient approximation. 

It is clear that the optimisation problem with $\zeta^d$ is simpler than those with $\zeta^d_J$. In this paper, we choose the simpler set $\zeta^d$ mainly for computational stability and efficiency.

Similarly, one can define $\psi^d_J$ in the partial data case following the same argument as for $\zeta^d_J$. The hierarchy of sets as well as the series of inequalities on the parameters with $\zeta^d_J$ are similar as in the complete data case.

\section{Observation Error Model} \label{sec:obserr} 

This section examines how our approach can be applied to the case of observation errors. We only describe this approach for the steady-state case, but the approach can be applied similarly to the transient case. 

Let us denote the moments of the measured numbers of molecules the \textit{measured moments} ($\boldsymbol{y}'_{:d'}$) in contrast with the \textit{true moments} ($\boldsymbol{y}_{:d}$). We assume an affine relationship between the true and measured moments:
\begin{equation} \label{eqn:obserrtransform}
\boldsymbol{y}'_{:d'} = B_{d',d} \boldsymbol{y}_{:d} + \boldsymbol{c}_{d'}
\end{equation}
where $B_{d',d} \in \mathbb{R}^{{N_d'} \times {N_d}}$ and $\boldsymbol{c}_{d'} \in \mathbb{R}^{N_{d'}}$. 

For example, consider a one-dimensional SRN with an independent additive Gaussian error such that
\begin{equation}
    X' = X + \epsilon, \ \epsilon \sim \text{N}(0,\sigma^2).
\end{equation}
Using the binomial expansion, one can then express each observed moment in terms of the true moments and the moments of the noise term. For instance, we have
\begin{equation}
    \begin{pmatrix}
    E(X'^0) \\
    E(X'^1) \\
    E(X'^2)
    \end{pmatrix} = \begin{pmatrix}
    1 & 0 & 0\\
    E(\epsilon) & 1 & 0\\
    E(\epsilon^2) & 2E(\epsilon) & 1
    \end{pmatrix} \begin{pmatrix}
    E(X^0) \\
    E(X^1) \\
    E(X^2)
    \end{pmatrix},
\end{equation}
and thus, the error model satisfies the linear form \eqref{eqn:obserrtransform}.

Another common observation error model is the binomial model~\citep{tang2020baynorm}, where each molecule is detected with probability $p$:
\begin{equation}
    X' | X \sim \text{Binomial}(X,p).
\end{equation}
This model also satisfies the linear error model \eqref{eqn:obserrtransform}, and the resulting matrix $B_{d',d}$ depends explicitly on $p$.

Defining $\boldsymbol{z}'_j = k_j \boldsymbol{y}'_{:d'}$ and defining the matrix form as $Z'\in\mathbb{R}_+^{N_{d'}\times R}$, we can write
\begin{equation}
    Z' = B_{d',d} Z + C_{d'},
\end{equation}
where the $j$-th column in $C_{d'}\in\mathbb{R}_+^{N_{d'}\times R}$ is $k_j \boldsymbol{c}_{d'}$.

Under the assumption that the matrix $B_{d',d}$ and the vector $\boldsymbol{c}_{d'}$ are known or can be estimated given the error model, one can form sets as described in the previous sections to constrain the parameters:
\begin{widetext}
\begin{equation}
	\zeta^{d',d}  = \left\{{\setstretch{1.314} \begin{array}{l}\boldsymbol{k} \in \mathcal{K}\\ \boldsymbol{y}_{:d} \in {\mathbb{R}_+^{N_d}} \\ Z \in  \mathbb{R}_+^{N_d\times R} \\ \boldsymbol{y'}_{:d'} \in {\mathbb{R}_+^{N_{d'}}} \\ Z' \in \mathbb{R}_+^{N_{d'}\times R} \end{array}} {\setstretch{1.314} \begin{array}{|ll}
        \boldsymbol{y}'_{:d'} = B_{d',d} \boldsymbol{y}_{:d} + \boldsymbol{c}_{d'}\\
        Z' = B_{d',d} Z + C_{d'} \\
		k_j \hat{\boldsymbol{y}}'_{:d',l} \leq \boldsymbol{z}'_j \leq k_j \hat{\boldsymbol{y}}'_{:d',u}, & j = 1,\ldots, R\\
			\text{trace}(A^d_{\boldsymbol{\alpha}} Z) = 0, & |{\boldsymbol{\alpha}}| \leq d - d_b + 1\\			\boldsymbol{z}_j^\top\boldsymbol{h}_{:N_d} = k_j, & j = 1,\ldots, R\\
			M_d^s(\boldsymbol{y}_{:d}) \succeq 0, & s = 0, \ldots, S\\ 
   ZC^\top = \boldsymbol{y}_{:d}\boldsymbol{c}^\top\\
	\end{array}}\right \},
\end{equation}
\end{widetext}
where $\hat{\boldsymbol{y}}'_{:d',l},\hat{\boldsymbol{y}}'_{:d',u}$ are element-wise intervals on the observed moments. 

Note that if the distribution of the error model is unknown, the relevant moments or parameters in $B_{d',d}$ and $\boldsymbol{c}_{d'}$ multiply $Z$ and thus lead to a non-linear, non-convex optimisation problem.

\section*{References}

\bibliography{ref.bib}

\end{document}